\documentclass{PoS}
\usepackage{graphicx}
\usepackage{bm}
\usepackage{amssymb}
\usepackage{amsmath}
\usepackage{hep}
\usepackage{multirow}

\title{Neutral Higgs-pair Production at one-loop from a Generic 2HDM}

\ShortTitle{Neutral Higgs-pair Production at one-loop from a Generic 2HDM}

\author{\speaker{David L\'opez-Val} \\
        High Energy Physics Group, Dept. ECM and Institut de Ci\`encies del Cosmos \\
        Universitat de Barcelona\\ Av. Diagonal 647, E-08028 Barcelona, Catalonia, Spain\\
        E-mail: \email{dlopez@ecm.ub.es}}

\author{Joan Sol\`a \\
        High Energy Physics Group, Dept. ECM and Institut de Ci\`encies del Cosmos \\
        Universitat de Barcelona\\ Av. Diagonal 647, E-08028 Barcelona, Catalonia, Spain\\
        E-mail: \email{sola@ecm.ub.es}}

\abstract{We present a one-loop analysis of the pairwise production of neutral
Higgs bosons ($\hzero\Azero, \Hzero\Azero$) at linear colliders, 
such as the ILC and CLIC, within the general Two-Higgs-Doublet Model (2HDM). 
We single out sizable radiative corrections, which can well reach the level 
of $|\delta\,\sigma|/\sigma \sim 50 \%$ and may be either positive 
(typically for $\sqrt{s} \simeq 0.5\,\TeV$) and negative (for $\sqrt{s} \gtrsim 1\,\TeV$).
These large quantum effects, obtained in full agreement with the current phenomenological bounds and the stringent theoretical constraints on the 
parameter space of the model,
can be traced back to the enhancement capabilities
of the triple-Higgs self-interactions -- a trademark feature of the 2HDM, with no counterpart\
in e.g. the Minimal Supersymmetric Standard Model. 
In the most favorable scenarios, the Higgs-pair cross sections may be
boosted up to barely 30 fb at the fiducial
center-of-mass energy of 500 GeV -- amounting to  $\mathcal{O}(\sim 10^3)$ events per 
500 $\invfb$ of integrated luminosity. We also compare these results with several complementary
double and triple Higgs-boson production mechanisms at order $\mathcal{O}(\alpha^3_{ew})$ and leading $\mathcal{O}(\alpha^4_{ew})$,
and we spotlight a plethora of potentially distinctive signatures of a Two-Higgs-Doublet structure of non-supersymmetric nature.}

\FullConference{RADCOR 2009 - 9th International Symposium on Radiative Corrections 
(Applications of Quantum Field Theory to Phenomenology) \\
                October 25-30 2009\\
                Ascona, Switzerland}

\newcommand{\eeAl}{\HepProcess{\APelectron\Pelectron \HepTo \PHiggspszero \PHiggslightzero}}

\newcommand{\complet}{\HepProcess{\APelectron\Pelectron \HepTo \PHiggspszero \PHiggslightzero/\Azero\Hzero}}

\newcommand{\retildehat}{\ensuremath{\mbox{Re}\,\hat{\Sigma}}}

\newcommand{\hzero}{\ensuremath{\PHiggslightzero}} 
\newcommand{\Hzero}{\ensuremath{\PHiggsheavyzero}} 
\newcommand{\Azero}{\ensuremath{\PHiggspszero}} 

\newcommand{\Hpm}{\ensuremath{\PHiggspm}} 
\newcommand{\bsg}{\ensuremath{\mathcal{B}(b \to s \gamma)}}
\newcommand{\CP}{\ensuremath{CP}}
\newcommand{\jump}{\vspace{0.2cm}}

\begin{document}

\section{Introduction}
\label{sec:intro}

The Two-Higgs-Doublet Model (2HDM) 
is a particularly simple extension of the Standard Model (SM)
which already encompasses outstanding new phenomenology \cite{hunter}.
In addition, a Two-Higgs-Doublet structure naturally
emerges as the low-energy realization of some more fundamental theories
-- viz. the Higgs sector of the Minimal Supersymmetric Standard Model (MSSM) \cite{mssm}. 
The 2HDM spectrum contains two neutral \CP-even ($\hzero, \Hzero$), one
\CP-odd ($\Azero$) and two charged Higgs bosons ($\Hpm$) and, as a most
distinctive feature, it allows 
triple (3H) and quartic (4H) Higgs self-interactions
to be largely enhanced -- in contrast to the MSSM, wherein such couplings
are restrained by the gauge symmetry. 
The phenomenological impact of such potentially large 3H self-interactions
has been actively investigated at linear colliders within a manifold of
processes, to wit: the tree-level production of triple Higgs-boson final
states \cite{giancarlo}; the double Higgs-strahlung channels $hh\PZ^0$ \cite{arhrib}; and
the inclusive Higgs-pair production via gauge-boson fusion \cite{neil}.
Likewise, the $\Pphoton\Pphoton$ mode of linac facilities has also been considered,   
in particular within the loop-induced 
single \cite{nicolas} and double Higgs production processes \cite{twophoton}. 
All these mechanisms would be capable to yield large Higgs production rates
and furnish experimental signatures which, 
owing to the clean environment inherent to linear colliders,
might enable for precise measurements of e.g. 
the Higgs boson masses, their couplings to fermions and gauge bosons
and their own self-interactions; this means, to reconstruct
the Higgs potential itself. Moreover, these large rates could be also revealing by themselves,
as they could not be deemed e.g. to a SUSY origin due to the intrinsically different nature of
Higgs self-interactions. 

\jump
Similary, double Higgs boson (2H) production may be also instrumental
at future linac facilities \cite{Djouadi:1992pu}. 
Such processes cannot proceed at the tree-level in the SM, which means that, if we
would detect a sizable rate of 2H final states e.g. of 
the sort  $\APelectron\Pelectron\to
\hzero\Azero; \Hzero\Azero;\PHiggs^{+}\,\PHiggs^{-}$, this would entail an unmistakable sign of
new physics. Nonetheless, a tree-level analysis of these processes
is most likely insufficient to disentangle e.g. 
SUSY and non-SUSY extensions of the Higgs sector, as the tree-level $h \Azero\PZ^0$ 
couplings ($h = \hzero, \Hzero$) are purely gauge, and hence both models can give rise to similar 
cross-sections at the leading-order. One-loop corrections to these 2H production rates 
are therefore required to this effect. A number of studies are available
within the MSSM \cite{mssmloop} whilst,
on the 2HDM side, the efforts were first concentrated
on the production of charged Higgs pairs \cite{charged_2hdm} and, only very recently, 
they have been extended to the neutral sector \cite{main}.


\section{Computation setup}
\label{sec:setup}

The general 2HDM\,\cite{hunter} is obtained upon
canonical extension of the SM Higgs sector with a second $SU_L(2)$
doublet with weak hypercharge $Y=+1$. 
The seven free parameters $\lambda_i$ in the general, \CP-conserving,
2HDM can be sorted out as follows: the masses of
the physical Higgs particles ($M_{h^0}$, $M_{H^0}$, $M_{A^0}$,
$M_{H^\pm}$), $\tan \beta$ (the ratio of the two VEV's $\langle
H_i^0\rangle$), 
the mixing angle $\alpha$ between the two $\CP$-even states; and
one Higgs self-coupling $\lambda_5$. Additionally, the Higgs couplings
to fermions must be engineered so that no tree-level flavor changing
neutral currents (FCNC) are allowed: this gives rise to the following
basic scenarios, namely Types-I and II 2HDM -- see \cite{hunter} for details. 
Further constraints must be imposed to assess that the SM behavior
is sufficiently well reproduced up to the energies explored so
far, namely: $i)$ the perturbativity and unitarity bounds \cite{unitarity}; $ii)$
the approximate $SU(2)$ custodial symmetry, which requires
$|\delta\rho_{2HDM}|\lesssim 10^{-3}$ \cite{custodial}; and $iii)$
the consistency with the low-energy radiative $B$-meson decays (which
demands $M_{H^{\pm}}
\gtrsim 300$ GeV for $\tan \beta \ge 1$
in the case of type-II 2HDM \cite{bound_charged}). We refer the reader
to \cite{main} for further details on the model setup and
the constraints.

\jump
Hereafter, our endeavor will be basically threefold:
i) to single out the regions in the 2HDM parameter 
space for which the $\complet$ production is optimized;
ii) to quantify the numerical impact of the associated quantum effects; and
iii) to correlate the latter with the enhancement capabilities of 3H
self-interactions. For definiteness, let us concentrate on the 
$\hzero\Azero$ channel. Our starting point shall be the $\APelectron\Pelectron \to \hzero\Azero$
scattering amplitude at 1-loop, 
\begin{equation}
{\cal M}_{\eeAl} = \sqrt{\hat{Z}_{\hzero}}\, \,{\cal
M}_{\eeAl}^{(0)}
 + {\cal M}^{(1)}_{\eeAl} \nonumber  + \delta\,{\cal
 M}^{(1)}_{\eeAl} \label{eq:amp} \, , 
\end{equation}
\noindent where $\sqrt{\hat{Z}_{\hzero\hzero}}$ stands for the finite WF renormalization
of the external $\hzero$ leg, which we shall expand as 
$\sqrt{\hat{Z}_{\hzero\hzero}} = 1 - \frac{1}{2}\,\retildehat'_{\hzero\hzero}(M^2_{\hzero})$,
in order to retain only those contributions which are at leading-order in the
triple-Higgs self-couplings. In turn, ${\cal M}^{(1)}_{\eeAl}$ includes the entire
set of $\mathcal{O}(\alpha^2_{ew}, \alpha_{ew}\,\alpha_{em})$ one-loop contributions,
to wit: i) the vacuum polarization corrections to the $\PZ^0$-boson propagator and the 
$\PZ^0-\Pphoton$ mixing; ii) the loop-induced $\Pphoton\Azero\hzero$ interaction;
iii) the vertex corrections to the $\APelectron\Pelectron\PZ^0$ and $\hzero\Azero\PZ^0$
interactions; and iv) box-type contributions. Finally, the 1-loop counterterm $\delta\,{\cal
 M}^{(1)}_{\eeAl}$ guarantees that the overall amplitude $\mathcal{M}$ is UV-finite.
Such counterterm ought to be anchored by a set of renormalization conditions;
for the SM fields and parameters, we stick to the conventional on-shell scheme in the Feynman gauge,
see e.g. \cite{renorm_sm}. Concerning the renormalization
of the 2HDM Higgs sector, we also employ an on-shell scheme
whose implementation is discussed in full detail in ~\cite{main}.

\section{Numerical results}
\label{sec:numerical}

The basic quantities of interest are: i) the
predicted cross section at the Born-level
$\sigma^{(0)}$ and at 1-loop $\sigma^{(0+1)}$, in which we include
the full set of $\mathcal{O}(\alpha^3_{ew})$ corrections, and also the leading
$\mathcal{O}(\alpha^4_{ew})$ which come from the squared of the scattering amplitude $\mathcal{M}$; and ii) the
relative size of the 1-loop radiative corrections, which we track
through the parameter $\delta_r = \frac{\sigma^{(0+1)}-\sigma^{(0)}}{\sigma^{(0)}} $.
Throughout the present work
we make use of the standard computational packages \cite{feynarts}.
For definiteness, we sort out the Higgs boson masses as follows:
\begin{center} \footnotesize{
\begin{tabular}{ccccc}
 & $M_{\PHiggslightzero}\,[\GeV]$ & $M_{\PHiggsheavyzero}\,[\GeV]$ & $M_{\PHiggspszero}\,[\GeV]$ & $M_{\PHiggs^\pm}\,[\GeV]$ \\ \hline
Set A & 130 & 150 & 200 & 160 \\ 
Set B & 150 & 200 & 260 & 300 
\end{tabular} }
\end{center}
We shall use type-I 2HDM for Set A, and type-II for Set B -- as, in the latter case, $M_{\PHiggs^{\pm}}$
is suitable to elude $\bsg$ constraints.
The predictions for both sets are quoted in Table~\ref{tab:res} for different
values of the $h\Azero\PZ^0$ tree-level coupling and maximally enhanced
3H self interactions -- namely $\tan\beta = 1$ and $\lambda_5 \simeq -10$ (resp. $-8$)
for Set A (resp. Set B).

\begin{table}[htb]
\begin{center}
\footnotesize{\begin{tabular}{|c|c|ccc|ccc|}
\hline
\multicolumn{2}{|c|}{\,}  & \multicolumn{3}{|c|}{$\hzero\Azero$} 
& \multicolumn{3}{|c|}{$\Hzero\Azero$} \\ \hline
\multicolumn{2}{|c|}{$\sqrt{s} = 0.5\,\TeV$} & $\alpha = \beta$ & $\alpha = \beta - \pi/6$ 
& $\alpha = \pi/2$
 &$\alpha = \beta-\pi/2$ & $\alpha = \beta - \pi/3$ 
& $\alpha = 0$
\\ \hline \hline
\multirow{2}{1cm}{Set A} & $\sigma_{max}\,[\femtobarn]$ & 26.71 &
 20.05 & 13.10 &  
26.32 & 17.53 & 10.29 \\ 
& $\delta_r\,[\%]$ & 31.32 & 31.42 & 28.81   & 48.45 & 31.82 & 16.10\\ \hline
 \multirow{2}{1cm}{Set B} & $\sigma_{max}\,[\femtobarn]$ & 11.63 
  & 9.08 & 6.36 & 5.00 & 3.28 & 1.95  \\
 & $\delta_r\,[\%]$ & 35.17 & 40.68 & 47.86  & 71.81 & 50.22 & 34.15 \\ \hline 
\multicolumn{1}{c}{\,}  \\ 
%
\hline
\multicolumn{2}{|c|}{$\sqrt{s} = 1\,\TeV$} &$\alpha = \beta$ & $\alpha = \beta - \pi/6$ 
& $\alpha = \pi/2$
 &$\alpha = \beta-\pi/2$ & $\alpha = \beta - \pi/3$ 
& $\alpha = 0$
\\ \hline \hline
\multirow{2}{1cm}{Set A} & $\sigma_{max}\,[\femtobarn]$ 
& 4.08 & 2.70 & 1.56 & 3.59 & 2.52 & 1.55   \\ 
 & $\delta_r\,[\%]$ & -58.42 & -63.28 & -68.11 & -62.62  & -65.09 & -67.75 \\ \hline 
\multirow{2}{1cm}{Set B} & $\sigma_{max}\,[\femtobarn]$ 
& 6.11 & 4.22 & 2.86 & 5.17 & 3.98 & 2.75  \\ 
 & $\delta_r\,[\%]$ & -30.16 & -35.62 & -34.58 & -36.80  & -35.14 & -32.80  \\ \hline  
\end{tabular}}
\end{center}
\caption{\footnotesize{Maximum total cross section $\sigma^{(0+1)}(\eeAl)$ 
and relative radiative correction ($\delta_r$) at
$\sqrt{s} = 0.5\,\TeV$ and $\sqrt{s} = 1\,\TeV$,
for Sets A and B of Higgs masses. The results are obtained at fixed $\tan\beta
= 1$ and different values of $\alpha$, with $\lambda_5 \simeq -10$ (resp. $-8$)
for Set A (resp. Set B).}}
\label{tab:res}
\end{table}
\begin{figure}[htb]
\begin{center}
\begin{tabular}{ccc}
 \hspace{-0.6cm}\resizebox{!}{6.1 cm}{\includegraphics{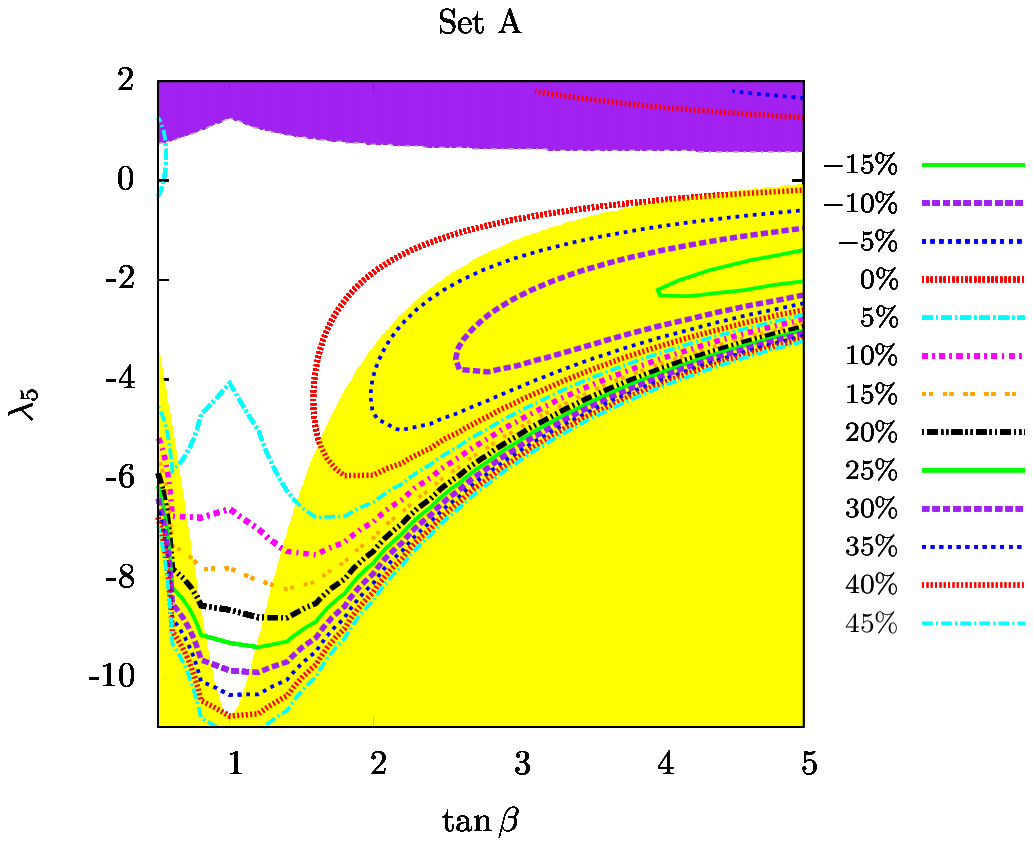}} & \hspace{0.5cm} &
 \hspace{ -0.6cm}\resizebox{!}{5.7 cm}{\includegraphics{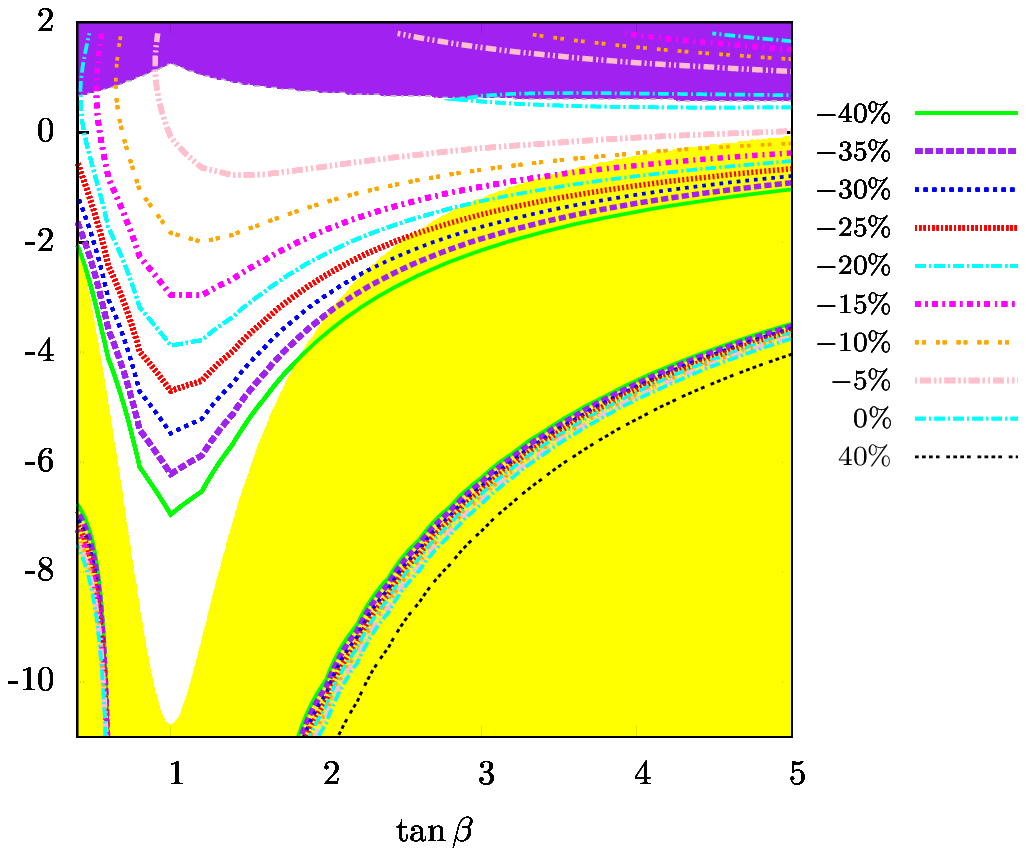}} \\
\end{tabular}
\caption{\footnotesize{Contour plots of the quantum corrections $\delta_r$ (in \%) 
to the $\eeAl$ cross-section as
a function of $\tan\beta$ and $\lambda_5$, for Set A of Higgs boson
masses, $\alpha = \beta$ and $\sqrt{s}
= 0.5\,\TeV$ (left panel), $\sqrt{s} = 1.0\,\TeV$ (right panel).
The shaded areas are excluded by the vacuum stability (top purple area)
and the unitarity bounds (bottom yellow area).}  \label{fig:scan}}
\end{center}
\end{figure}

These results single out large quantum effects
($|\delta\sigma|/\sigma \sim 20-60\,\%$), which can be either positive (for
$\sqrt{s} \simeq 0.5\,\TeV$) or negative (
$\sqrt{s} \gtrsim 1\,\TeV$) and turn out to be fairly independent on the
details of the Higgs mass spectrum, the particular type of 2HDM and
the specific channel under analysis ($\hzero\Azero, \Hzero\Azero$). The corresponding
cross-sections at 1-loop lie in the approximate range of $2-30\,\femtobarn$
for $\sqrt{s} = 0.5\,\TeV$ -- up to barely $10^3-10^4$ events per
$500\,\invfb$. As for the radiative corrections themselves, in
Figure~\ref{fig:scan} we display their detailed behavior (for
$\hzero\Azero$) as a function of $\tan\beta$
and $\lambda_5$, as well as their interplay with the unitarity bounds (lower area, in yellow)
and the vacuum stability conditions (upper area, in purple). Notice that the former
disallows simultaneously large values of $\tan\beta$ and $\lambda_5$, whereas
the latter enforces $\lambda_5 \lesssim 0$. 
The largest attainable quantum effects (positive or negative, depending
on $\sqrt{s}$) are identified
in a valley-shaped region at $\tan\beta \simeq 1$ and $|\lambda_5| \sim 5-10$.
In such regimes, a subset of 3H self-couplings becomes substantially augmented
-- their strenght growing with $\sim |\lambda_5|$ -- and 
stand as a preeminent source of radiative corrections -- via Higgs-boson
mediated 1-loop corrections to the $h\Azero\PZ^0$ vertex. As a result the latter
interaction, which is purely gauge at the tree-level, becomes drastically
promoted at the 1-loop order. Upon
simple power counting arguments and educated guess
we may barely estimate the loop-corrected $\hzero\Azero\PZ^0$ coupling as 
$ \Gamma^{eff}_{\hzero\Azero\PZ^0} \sim
\Gamma^{0}_{\hzero\Azero\PZ^0}\lambda^2_{3H}/16\pi^2\,s\,\times\,f(M^2_{\hzero}/s,
M^2_{\Azero}/s)$, where $f(M^2_{\hzero}/s, M^2_{\Azero}/s)$ is a dimensionless form
factor. 

\begin{figure}[htb]
\begin{center}
\begin{tabular}{cc}
\hspace{-1.cm} \includegraphics[scale=0.52]{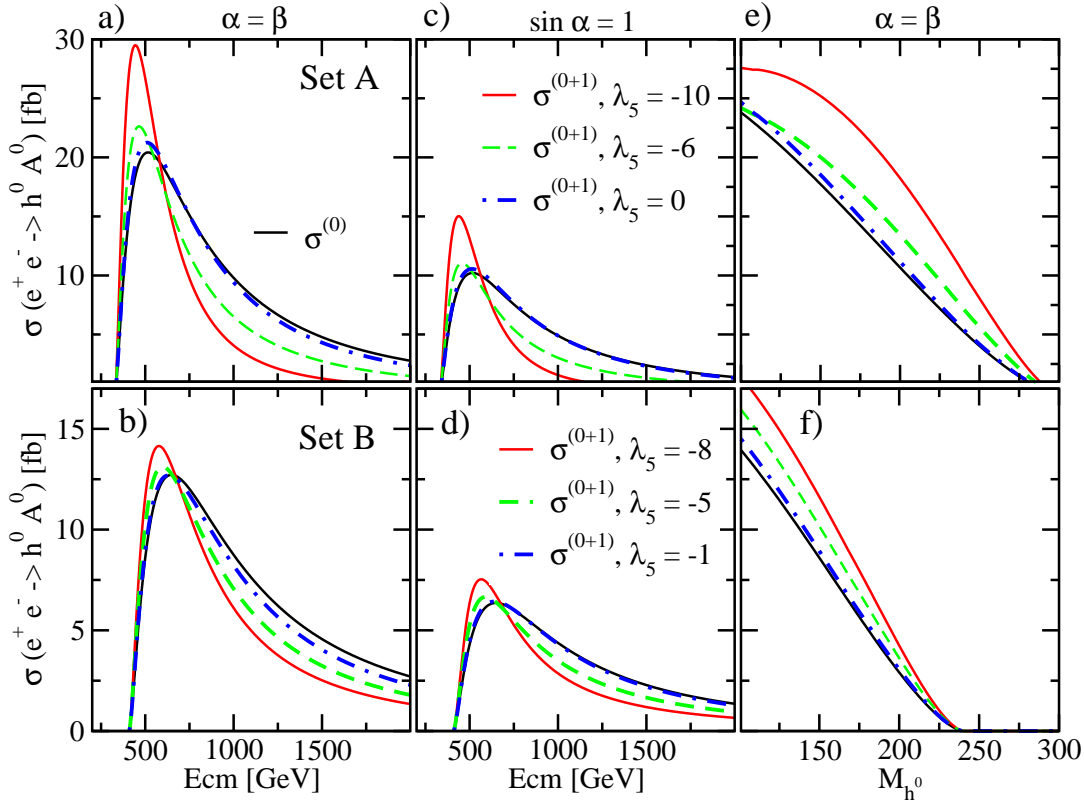}
\end{tabular}
\caption{\footnotesize{Total cross section $\sigma(\eeAl)$ (in \femtobarn) at the
tree-level and at one-loop for different values of $\lambda_5$ and for Sets A (top panels)
and B (bottom panels) of Higgs boson masses. Displayed is the behavior of $\sigma$
as a function of $\sqrt{s}$ (left and center) and $M_{\hzero}$ (right panels).} \label{fig:over}}
\end{center}
\end{figure}

All these trademark phenomenological features become also patent in Figure~\ref{fig:over},
in which we explore the evolution of the $\hzero\Azero$
production cross-section for the following setups: a-b) as a function of $\sqrt{s}$,
for $\alpha = \beta$ (maximum $\hzero\Azero\PZ^0$ coupling); 
c-d) the same, for $\alpha = \pi/2$ (the so-called \emph{fermiophobic} limit
for the $\hzero$ boson within type-I 2HDM); and e-f) 
as a function of $M_{\hzero}$, for $\alpha = \beta$. 
The results are presented for Sets A (upper row) and B (lower row).
The phenomenological footprint of 3H self-couplings is easily read off
from the dramatic differences in the one-loop production rates when the value of
$\lambda_5$ is varied in the theoretically allowed range. 
As for the behavior with $M_{\hzero}$,
these plots illustrate a twofold effect on $\sigma$; 
one is purely kinematical, and accounts for the
decrease of the cross-section owing to the 
reduction of the available phase space; the other is dynamical,
and explains why the tree-level cross-sections decouple faster than
their corresponding one-loop counterparts; indeed, several 3H self-couplings
get boosted with $M_{\hzero}$ and can partially
counterbalance the phase space suppression.

\begin{table}
\begin{center}
\footnotesize{\begin{tabular}{|c|ccc|ccc|}  \hline
\multirow{2}{1cm}{} & $0.5\,{\rm TeV}$ &
$1.0\,{\rm TeV}$ & $1.5\,{\rm
TeV}$ & $0.5\,{\rm TeV}$ &
$1.0\,{\rm TeV}$ & $1.5\,{\rm
TeV}$
\\ \cline{2-7}
\multicolumn{1}{|c|}{\,} & \multicolumn{3}{|c|}{Set A} & \multicolumn{3}{|c|}{Set B} \\ \cline{1-7} 
$\sigma(\hzero\Azero)\,[\femtobarn]$ &  26.71 & 4.07 & 1.27 &  11.63 & 6.11 & 2.52 \\
$\sigma(\hzero\Hzero\Azero)\,[\femtobarn]$ & 0.02 & 5.03 & 3.55 & below thres. & 1.25 & 1.33 \\
$\sigma(\Hzero\PHiggs^+\PHiggs^-)\,[\femtobarn]$ & 0.17 & 11.93 & 8.39 & below thres. & 0.69 & 2.14 \\
$\sigma(\hzero\hzero + X)\,[\femtobarn]$ & 1.47 & 17.36 & 38.01 & 0.92 & 9.72 & 23.40 \\ \hline
\end{tabular}}
\end{center}
\caption{\footnotesize{Comparison of the predictions for the cross sections
corresponding to several Higgs-pair and triple-Higgs
production channels, for Sets A and B of Higgs masses;
$\tan\beta = 1$, $\alpha=\beta$, and three
different values of the center-of-mass energy. The
complementarity between the different channels at different
energies becomes manifest here.}}
\label{tab:compare}
\end{table}

\section{Discussion and conclusions}
\label{sec:conclusions}

We have undertaken a comprehensive study of the neutral 2H production from a generic 2HDM
in the context of the ILC/CLIC colliders. The upshot of our analysis spotlights very significant
radiative corrections, typically up to $|\delta\sigma/\sigma|\sim 50\%$, which may be either positive
(for $\sqrt{s} \simeq 0.5\,\TeV$) and negative (for $\sqrt{s} \simeq 1\,\TeV$ and above),
and basically concentrated in the regions with $\tan\beta \sim 1$ and $|\lambda_5| \sim 10$, with $\lambda_5 < 0$.
Such enlarged quantum effects can be basically traced back to the Higgs-mediated 1-loop
corrections to the $h\Azero\PZ^0$ interaction, these being sensitive to potentially
enhanced 3H self-couplings -- a genuine feature of the 2HDM, with no counterpart in the MSSM. 
In the most favorable scenarios, the predicted 2H rates furnish a few dozen fb
per $500\,\invfb$ of integrated luminosity. 
In practice, these 2H processes would boil down to 
$\Azero \to \APbottom\Pbottom / \tau^+\tau^-$, and $\hzero  \to \APbottom\Pbottom / \tau^+\tau^-$
or $ \hzero \to VV \to 4l,2l +$ mising energy, depending on the actual value of $M_{\hzero}$;
all these signatures being feasible in the clean environment of linac facilities.
Besides, the analysis of such pairwise Higgs events should be useful to enlighten
the inner structure of the underlying Higgs sector, in particular if combined with
complementary Higgs production channels such as $\APelectron\Pelectron \to hhh$ \cite{giancarlo}
and $\APelectron\Pelectron \to hh + X$ \cite{neil}. In Table~\ref{tab:compare} we quantify the
latter statement by plugging $\sigma^{(0+1)}$ for $\hzero\Azero$ together
with the predicted (leading-order) production rates for 
several complementary multi-Higgs channels in $\APelectron\Pelectron$ collisions,
assuming maximum 3H self-coupling enhancements (as e.g. Table~\ref{tab:res}). 
This pattern of signatures at different $\sqrt{s}$ is highly distinctive of the 2HDM and could not be
attributed to e.g. the MSSM,  
as they critically depend on the
3H self-couplings -- which are fully inconspicuous in the latter model owing to supersymmetric
invariance. Additional, and very
valuable, information can also be obtained from the study of the quantum
effects on the production of a neutral 2HDM Higgs boson in
association with the Z boson \cite{new}.

We conclude that 3H self-couplings may stamp genuine footprints
on multi-Higgs production processes, either at leading-order
or through quantum corrections, and 
provide a plethora of complementary signatures whose detection should be perfectly feasible
at future linac facilities and, if ever observed, might constitute a strong hint
of non-standard, non-supersymmetric Higgs physics.

\vspace{-0.15cm}
\paragraph{Acknowledgments} This work has been supported in part by the EU project RTN
MRTN-CT-2006-035505 Heptools. DLV acknowledges 
an ESR position of this
network and also the support
of the MEC FPU grant Ref. AP2006-00357. 
He also wishes to thank the hospitality of
the Theory Group at the Physikalisches Institut of the University of Bonn. 
JS has been supported in part by MEC and FEDER under project FPA2007-66665,
by the Spanish Consolider-Ingenio 2010 program CPAN CSD2007-00042
and by DIUE/CUR Generalitat de Catalunya under project 2009SGR502. 
%

\newcommand{\JHEP}[3]{ {JHEP} {#1} (#2)  {#3}}
\newcommand{\NPB}[3]{{ Nucl. Phys. } {\bf B#1} (#2)  {#3}}
\newcommand{\NPPS}[3]{{ Nucl. Phys. Proc. Supp. } {\bf #1} (#2)  {#3}}
\newcommand{\PRD}[3]{{ Phys. Rev. } {\bf D#1} (#2)   {#3}}
\newcommand{\PLB}[3]{{ Phys. Lett. } {\bf B#1} (#2)  {#3}}
\newcommand{\EPJ}[3]{{ Eur. Phys. J } {\bf C#1} (#2)  {#3}}
\newcommand{\PR}[3]{{ Phys. Rept. } {\bf #1} (#2)  {#3}}
\newcommand{\RMP}[3]{{ Rev. Mod. Phys. } {\bf #1} (#2)  {#3}}
\newcommand{\IJMP}[3]{{ Int. J. of Mod. Phys. } {\bf #1} (#2)  {#3}}
\newcommand{\PRL}[3]{{ Phys. Rev. Lett. } {\bf #1} (#2) {#3}}
\newcommand{\ZFP}[3]{{ Zeitsch. f. Physik } {\bf C#1} (#2)  {#3}}
\newcommand{\MPLA}[3]{{ Mod. Phys. Lett. } {\bf A#1} (#2) {#3}}
\newcommand{\JPG}[3]{{ J. Phys.} {\bf G#1} (#2)  {#3}}
\newcommand{\PTP}[3]{{\sl Prog. Theor. Phys. Suppl.} {\bf G#1} (#2)  {#3}}
\newcommand{\FP}[3]{{\sl Fortsch. Phys.} {\bf G#1} (#2)  {#3}}


\end{document}